\documentclass[amsmath,amssymb,aps,floatfix,prd,a4paper]{revtex4}

\usepackage{hyperref}
\usepackage[T1]{fontenc}
\usepackage{dcolumn}
\usepackage{bm}
\usepackage{epsfig}
\usepackage{amsfonts}
\usepackage{amssymb,amscd}
\usepackage{tabularx}
\usepackage{booktabs,multirow}
 \usepackage{amsmath} 

\def\lsim{\raise0.3ex\hbox{$<$\kern-0.75em\raise-1.1ex\hbox{$\sim$}}}

\def\gsim{\raise0.3ex\hbox{$>$\kern-0.75em\raise-1.1ex\hbox{$\sim$}}}

\newcommand{\be}{\begin{equation}}

\newcommand{\ee}{\end{equation}}

\def\beq{\begin{equation}}

\def\eeq{\end{equation}}

\def\beqa{\begin{eqnarray}}

\def\eeqa{\end{eqnarray}}

\newcommand{\ba}{\begin{eqnarray}}

\newcommand{\ea}{\end{eqnarray}}

\def\gappeq{\mathrel{\rlap {\raise.5ex\hbox{$>$}}

{\lower.5ex\hbox{$\sim$}}}}

\def\lappeq{\mathrel{\rlap{\raise.5ex\hbox{$<$}}

{\lower.5ex\hbox{$\sim$}}}}

\def\Toprel#1\over#2{\mathrel{\mathop{#2}\limits^{#1}}}

\begin{document}

\title{Inferring the neutral-current neutrino--nucleon cross section from IceCube data}

\author{Reinaldo {\sc Francener}}
\email{reinaldofrancener@gmail.com}
\affiliation{Institute of Physics and Mathematics, Federal University of Pelotas (UFPel), \\
  Postal Code 354,  96010-900, Pelotas, RS, Brazil. 
  \vspace{0.5cm}}

\author{Victor P. {\sc Gon\c{c}alves}}
\email{barros@ufpel.edu.br}
\affiliation{Institute of Physics and Mathematics, Federal University of Pelotas (UFPel), \\
  Postal Code 354,  96010-900, Pelotas, RS, Brazil. 
  \vspace{0.5cm}}

\begin{abstract}
\vspace{0.3cm}
Neutrinos interact only via the weak force, through the exchange of $W^{\pm}$ and $Z^{0}$ bosons in charged-current and neutral-current interactions, respectively. Recent experiments have measured the charged-current cross section at high energies, whereas neutral-current interactions remain largely unexplored experimentally. In this work, we investigate the sensitivity of the IceCube Neutrino Observatory to the neutral-current neutrino--nucleon cross section using 12 years of data. We consider two strategies to constrain this quantity: $(i)$ leveraging neutrino absorption and regeneration as neutrinos traverse the Earth; and $(ii)$ using the track-to-cascade event ratio. We find that current IceCube data can set only an upper limit on the neutral-current neutrino--nucleon cross section. We further show that IceCube-Gen2 should be able to constrain the neutral-current cross section with both methods considered here.
\end{abstract}


\maketitle

\vspace{1cm}

\section{Introduction}

The study of high- and ultra-high-energy extraterrestrial neutrino fluxes is among the primary goals of the IceCube Neutrino Observatory \cite{IceCube:2013low}. Their observation has important consequences for probing the Standard Model (SM) of particle physics, as well as for the search for beyond-SM signals through neutrino production, propagation, and interactions at the South Pole. Despite the extensive efforts of the IceCube Collaboration, only a few sources of these neutrinos have been identified to date, motivating the proposal of a next generation of neutrino observatories such as IceCube-Gen2 \cite{IceCube-Gen2:2020qha}, KM3NeT \cite{KM3Net:2016zxf}, Baikal-GVD \cite{Baikal-GVD:2018isr}, P-ONE \cite{P-ONE:2020ljt}, and TRIDENT \cite{TRIDENT:2022hql}.

At these energies, neutrinos typically interact via deep-inelastic scattering (DIS), mediated by $W^\pm$ and $Z^0$ exchange in charged-current (CC) and neutral-current (NC) processes, respectively. CC neutrino--nucleon cross sections have been explored by a variety of experiments, including accelerator neutrinos \cite{Mukhin:1979bd,Baranov:1978sx,Barish:1978pj,GargamelleNeutrinoPropane:1979kqo,deGroot:1978feq,Colley:1979rt,GargamelleSPS:1981hpd,Baker:1982ty,Berge:1987zw,Anikeev:1995dj,Seligman:1997fe,NuTeV:2005wsg,NOMAD:2007krq,MINOS:2009ugl,SciBooNE:2010slc,T2K:2013nor,ArgoNeuT:2014rlj,T2K:2014axs}, collider measurements \cite{FASER:2024hoe,FASER:2024ref,SNDLHC:2026oxu}, atmospheric neutrinos \cite{Super-Kamiokande:2026xsm}, and astrophysical neutrinos \cite{IceCube:2017roe,Bustamante:2017xuy,IceCube:2020rnc,Hajjar:2026uzs}. In contrast, the NC cross section remains much less constrained, with only a few extractions based on the NC-to-CC event ratio \cite{CHARM:1981zpm,CHARM:1980yym,CHARM:1983rtn,Abramowicz:1985xg,CCFR:1993urw,NuTeV:2001whx}. This scarcity of measurements reflects both the smaller NC cross section (relative to CC) and the experimental difficulty of identifying final states with an outgoing neutrino and disentangling them from CC backgrounds.

In this work, we investigate the prospects for constraining the NC neutrino--nucleon cross section using 12 years of the IceCube High-Energy Starting Events (HESE) sample \cite{IceCube:2023sov}. The IceCube Observatory can identify three distinct event topologies: tracks, cascades, and double cascades. NC interactions are reconstructed as cascades, along with CC interactions that do not produce a high-energy muon in the final state. This overlap between CC and NC signatures prevents event-by-event identification of NC interactions, allowing only a statistical inference of their contribution.  To infer the magnitude of the NC cross section, we employ two complementary approaches: $(i)$ leveraging the impact of the NC cross section on neutrino absorption and regeneration as neutrinos traverse the Earth before reaching the South Pole detector; and $(ii)$ exploiting the statistical contribution of NC interactions to the cascade topology. The analysis is performed considering two distinct treatments for the astrophysical neutrino flux. First, we consider the standard procedure, in which the normalization and spectral index are free parameters, which are fixed by the IceCube data, as well the NC cross  section. Additionally, we consider an optimistic scenario, in which the astrophysical flux is fixed in an independent analysis, and the IceCube data can be used to directly constrain the magnitude of the NC cross section. For both treatments of the astrophysical neutrino flux, we  estimate the impact of the modeling of its composition on our results. In particular,  we consider that the flux is produced in the astrophysical source by pion decay, which implies equal quantities of neutrinos of each flavor at Earth,  or that its composition is given by the muon dumped pion decay model.  As we will demonstrate below, current IceCube data can set only an upper limit on the neutral-current neutrino–nucleon cross section. However,  IceCube-Gen2 should be able to constrain the neutral-current cross section with both methods considered here.    

This paper is organized as follows. In the next section we review our methodology for cross section calculation, as well as to estimate events at neutrino observatories. In section \ref{sec:res} we present our results for limits on NC cross section through neutrino absorption and track/cascade ratio. The future sensitivity of IceCube-Gen2 to this observable is also estimated. Finally, in section \ref{sec:sum} we conclude with a brief summary.

\section{Formalism}
\label{sec:for}

In this section, we present a brief review the formalism and discuss the main ingredients used in our analysis to estimate the impact of neutral current interactions at IceCube events. In particular, we present the treatment of the neutrino - target cross section in the deep inelastic scattering regime and the calculation of the number of neutrino events at the IceCube Observatory.

\subsection{Neutrino-target cross section}

At high and ultra-high energies, the neutrino--nucleon cross section is dominated by deep-inelastic scattering (DIS). In this process, an incident neutrino with four-momentum $k=(E_\nu,0,0,E_\nu)$ exchanges a weak boson ($W^{\pm}$ or $Z^{0}$) with squared four-momentum transfer $q^{2}=-Q^{2}$, leaving the nucleon in a hadronic final state $X$. Throughout this work we restrict to the DIS region by requiring $Q>1.3\,\mathrm{GeV}$ and hadronic invariant mass $W>1.4\,\mathrm{GeV}$. In CC interactions, the incoming neutrino is converted into a charged lepton with energy $E_{\ell}$, whereas in NC interactions a neutrino of the same flavor remains in the final state. The DIS double-differential cross section for NC interactions is commonly expressed in terms of the Bjorken-$x$ and inelasticity $y$ variables as \cite{Paschos:2001np}
\begin{eqnarray}
\begin{aligned}
\frac{\mathrm{d}^{2}\sigma^{\nu(\bar \nu)}}{\mathrm{d} x\mathrm{d} y} = \frac{G^{2}_{F}m_{N}E_{\nu}}{\pi}\left(\frac{m^{2}_{Z^0}}{Q^{2}+m^{2}_{Z^0}} \right)^{2}\left\{    
xy^{2}F_{1}^{NC}(x,Q^{2})
+\left.  \left[ 1   - \left( 1+\frac{m_{N}x}{2E_{\nu}}\right)y  \right] F_{2}^{NC}(x,Q^{2}) 
\right.\right.    \\
\pm \left.  \left[ xy\left( 1 - \frac{y}{2} \right)   \right] F_{3}^{NC} (x,Q^{2}) \right\} \, ,
\end{aligned}
\label{eq:sigmaDISNC}
\end{eqnarray}
where $m_{Z^0}$ is the $Z^0$ mass and $F_i^{NC}$ are the NC nucleon structure functions. An analogous expression holds for CC scattering, with the appropriate replacements of gauge-boson mass, couplings, and structure functions \cite{Kretzer:2002fr}. In this work we use the CT18 \cite{Hou:2019efy,Yan:2022pzl} parton distribution functions (PDFs) to compute the structure functions entering both the NC and CC cross sections. {For simplicity, we do not account for PDF uncertainties propagated to the CC and NC cross sections, as these are approximately 2–3\% in the energy range of interest, which is significantly smaller than the uncertainties associated with the CC cross section extracted from IceCube data. }


In our analysis, in addition to CC and NC neutrino--nucleon interactions, we also consider the Glashow resonance \cite{Glashow:1960zz}. This process is characterized by the annihilation of an electron antineutrino with an electron, producing an on-shell $W^{-}$ boson. One event candidate has been observed at IceCube so far \cite{IceCube:2021rpz}, and the corresponding cross section is dominant for incident (anti)neutrino energies near 6.3~PeV.

\subsection{Neutrino events at the IceCube Observatory}

The main observable at the IceCube Neutrino Observatory is the number of neutrino initiated events, which is given by
\begin{eqnarray}
\mathrm{d}N_\mathrm{events} = 
T \sum_{\alpha, l}
N_{\mathrm{effe},\alpha}(E_{\nu})\times 
\Phi_{\nu_l} (E_{\nu})\times
\sigma _{\nu _l \alpha}(E_\nu) \times
T_{\nu_l}(E_\nu , \theta_{z})\,
{\mathrm{d}(E_\mathrm{dep})\, \mathrm{d}\Omega}\, ,
\label{eq:eventos}
\end{eqnarray}
where $T$ is the data-taking exposure time, $N_{\mathrm{effe}, \alpha}$ is the effective number of targets of type $\alpha$, $\Phi_{\nu_l}$ is the incident neutrino flux, $\sigma_{\nu_l \alpha}(E_\nu)$ is the neutrino--target cross section, and $T_{\nu_l}(E_\nu,\theta_z)$ is the transmission coefficient for neutrinos of flavor $l=e,\mu,\tau$ traversing the Earth at zenith angle $\theta_z$. For $N_{\mathrm{effe},\alpha}$ we use the results of Ref.~\cite{Palomares-Ruiz:2015mka}, where this quantity is provided as a function of the electromagnetic-equivalent deposited energy $E_\mathrm{dep}$, valid for all processes and topologies. The transmission coefficient $T_{\nu_l}(E_\nu,\theta_z)$ is computed following Refs.~\cite{Palomares-Ruiz:2015mka,Rakshit:2006yi,Francener:2024bfm} by solving the coupled neutrino-transport equations in matter, including neutrino regeneration from neutral-current interactions and $\tau$-lepton decay. We assume the Earth density profile from the Preliminary Reference Earth Model (PREM) \cite{Dziewonski:1981xy}.

Finally, the last quantity of Eq.~\ref{eq:eventos} is the astrophysical neutrino flux that reaches the neutrino observatory. In our analysis, we are parameterizing this flux as a power law of the form
\begin{eqnarray}
\Phi_{\nu_l} (E_\nu) = f_l\, \Phi_0 \left( \dfrac{E_\nu}{ 100\;\mathrm{TeV}}\right) ^{-\gamma} \; 10^{-18} \mathrm{GeV}^{-1} \mathrm{s}^{-1} \mathrm{sr}^{-1}\mathrm{cm}^{-2}\, ,
\label{eq:fluxo}
\end{eqnarray}
with $\Phi_0$ and $\gamma$ being its normalization and spectral index and $f_l$ the flavor ratio of neutrinos with flavor $l = e, \mu, \tau$. Recently, the IceCube Collaboration reported evidence for the astrophysical neutrino flux given by a broken power law \cite{IceCube:2025tgp,IceCube:2025ewu}, with $E_{\mathrm{cut}} \approx 25$~TeV being the neutrino energy between two spectral indexes. As we are considering in our analysis events with $E_\mathrm{dep} >60$~TeV, we will not consider this scenario for the neutrino flux. In addition, we consider two cases for the composition of the neutrino flux: $(i)$ flux produced in the astrophysical source by pion decay (PD), which produces approximately equal quantities of neutrinos of each flavor at Earth after decoherence effects ($f_e \approx 0.98$, $f_\mu \approx 1.03$ and $f_\tau \approx 0.99$); and $(ii)$ muon dumped pion decay (MDPD), which predicts $f_e \approx 0.66$, $f_\mu \approx 1.21$ and $f_\tau \approx 1.13$ for the composition fraction of the astrophysical neutrino flux. These values were obtained using the best fit from NuFIT-6.0 \cite{Esteban:2024eli}. We will not consider here the scenario of neutrinos from neutron decay, given that is disfavored by current IceCube MESE analysis \cite{IceCube:2025ole}.

\section{Results}
\label{sec:res}

In what follows, we present constraints on the neutral-current  neutrino–nucleon cross section using 12 years of IceCube HESE data, derived considering two complementary strategies. In subsection \ref{sub:absorption}, we first investigate the feasibility of constraining the NC cross section by analyzing its impact on neutrino absorption during propagation through the Earth. Conversely, in subsection \ref{sub:cascade}, we investigate the impact of neutral current interactions on cascade-like events to estimate the constraints on the NC cross section achievable with current IceCube data. In contrast to the analyses performed in subsections \ref{sub:absorption} and \ref{sub:cascade}, where both the normalization and spectral index of the astrophysical neutrino flux are fitted, subsection \ref{sub:optimistic} explores an optimistic scenario assuming a fixed neutrino flux to derive the corresponding constraints. Finally, subsection \ref{sub:gen2} presents the prospects for IceCube-Gen2.

\subsection{Constraining the NC cross section using Earth-absorption effects}
\label{sub:absorption}

\begin{figure}[t]
	\centering
	\begin{tabular}{ccccc}
\includegraphics[width=0.5\textwidth]{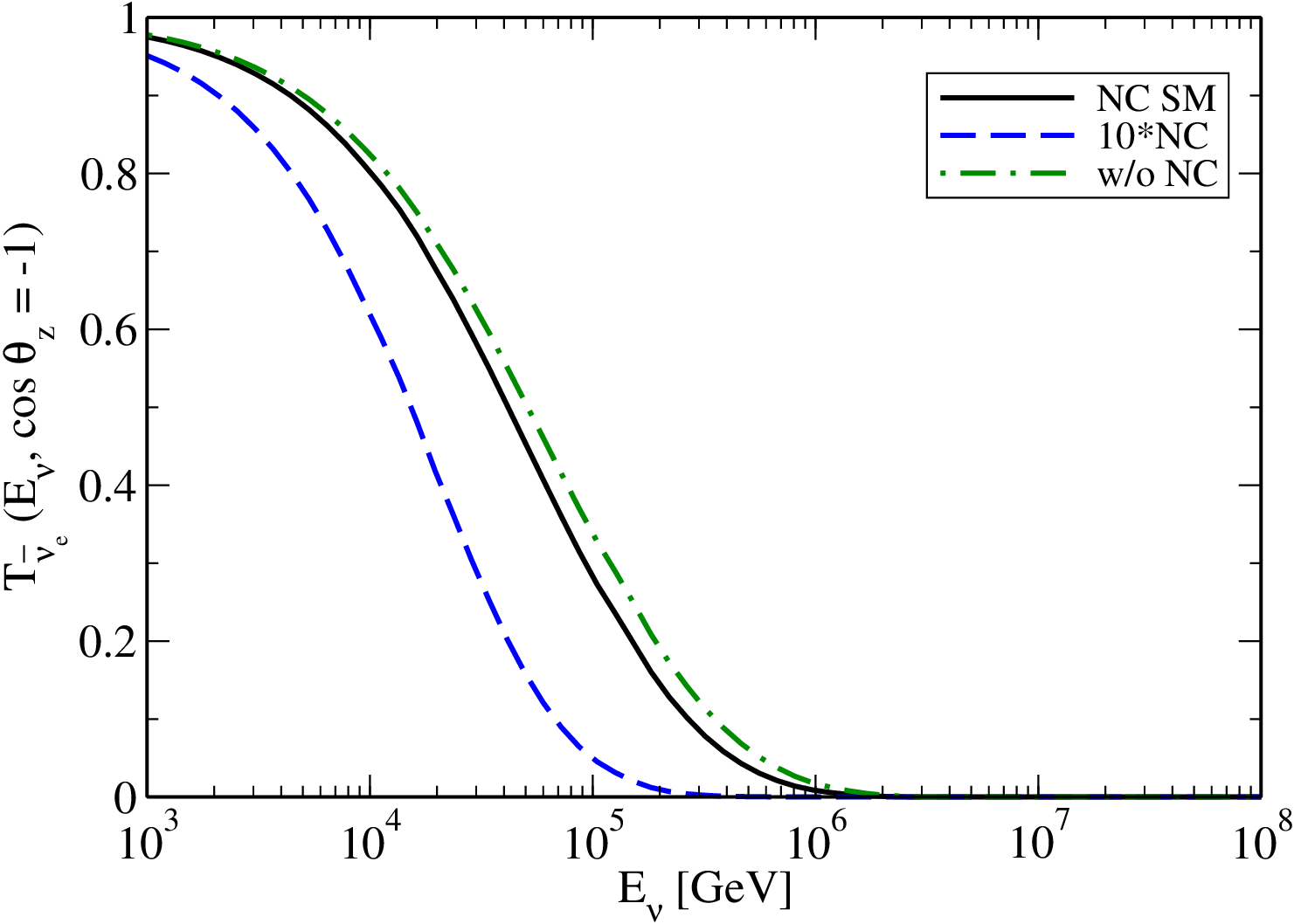}  &
\includegraphics[width=0.5\textwidth]{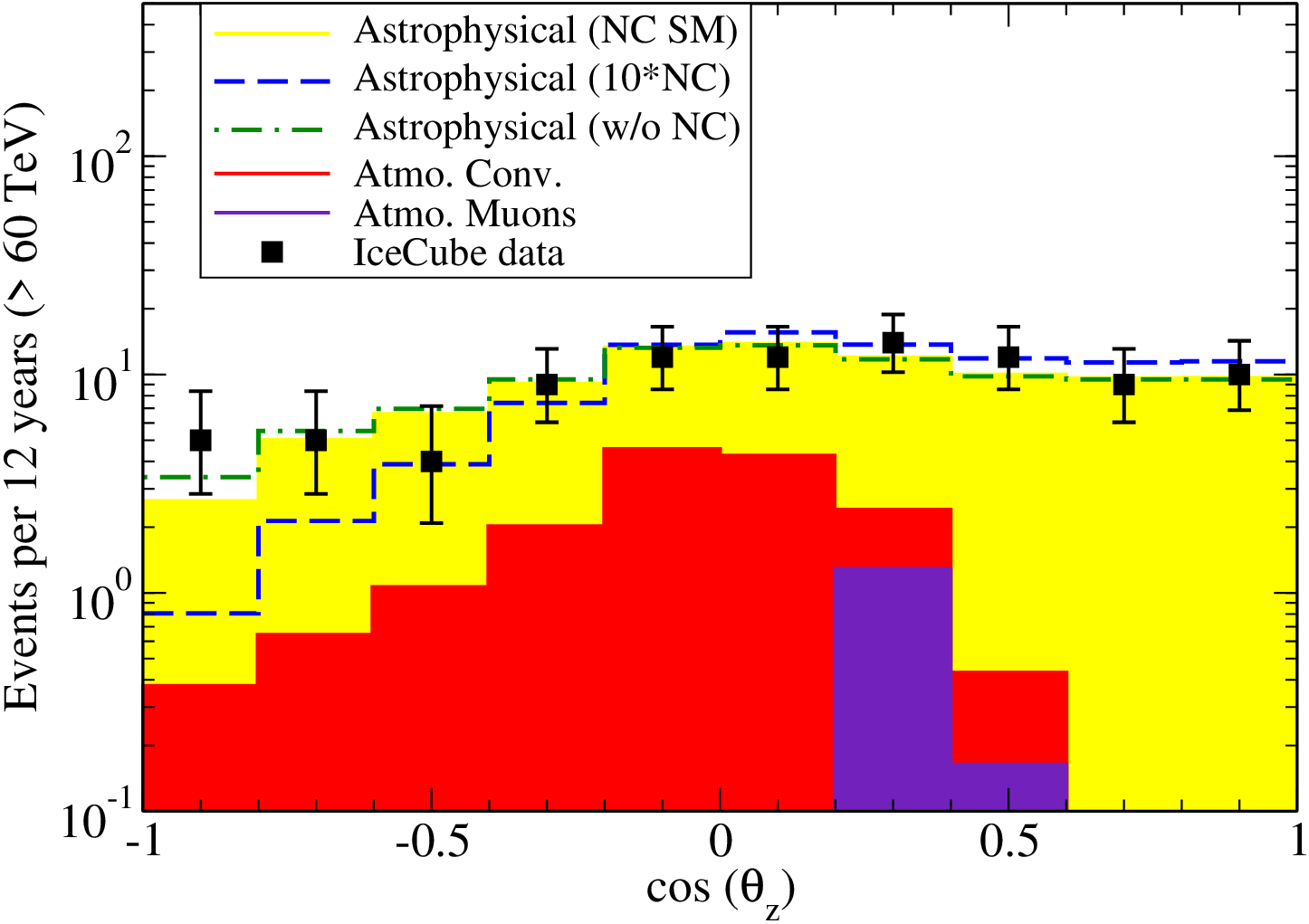}
	\end{tabular}
\caption{ {\bf Left panel:} Transmission coefficient of electron antineutrinos versus energy for three representative neutral-current cross-section normalizations. Results are shown for $\mathrm{cos},\theta_z = -1$, assuming a power-law neutrino flux with $\gamma = 3.0$. {\bf Right panel:} Distribution of HESE events per $\cos\theta_z$ bin, based on 12 years of data, comparing different normalizations of the NC cross section. }
\label{fig:Transmission}
\end{figure}

The astrophysical neutrino flux is typically parameterized as the power law in Eq.~\ref{eq:fluxo}. Since only a small number of sources have been identified so far, it is commonly modeled as isotropic. At TeV energies and above, the Earth is no longer transparent to neutrinos, and the transmission coefficient depends on the incident neutrino energy, the zenith angle, and the neutrino flavor (see, e.g., Refs.~\cite{Palomares-Ruiz:2015mka,Rakshit:2006yi,Francener:2024bfm}). In Fig.~\ref{fig:Transmission} (left panel) we show the transmission coefficient for electron antineutrinos as a function of energy, assuming that neutrinos traverse the full Earth before reaching the IceCube Observatory ($\mathrm{cos}\,\theta_z=-1$) and taking $\gamma=3.0$. The results are obtained by solving the coupled differential equations, including regeneration via neutral-current interactions and $\tau$ decay~\cite{Francener:2024bfm}. We explore three NC scenarios: the Standard Model (SM) prediction (solid black line), a tenfold enhancement of the cross section (dashed blue), and the absence of any NC contribution (dot-dashed green). As expected, our results indicate that neutrino absorption is dominated by CC interactions, which convert neutrinos into their corresponding charged leptons. However, they also indicates that the contribution of NC interactions increases with the energy. Moreover, the transmission coefficient is strongly modified if new effects, associated e.g. with beyond the SM physics, increase the magnitude of NC cross section.

In Fig.~\ref{fig:Transmission} (right panel) we compare the IceCube HESE events with deposited energy $>60\,\mathrm{GeV}$ as a function of $\mathrm{cos}\,\theta_z$ with our predictions for three choices of the NC cross-section normalization. These three cases correspond to the best fits obtained with a maximum-likelihood analysis of events binned in deposited energy, where the flux normalization and spectral index are treated as free parameters. For the central (SM) NC prediction, assuming the pion
decay (PD) composition at Earth, we find $\Phi_0=1.31$ and $\gamma=2.93$. The atmospheric neutrino and muon backgrounds are taken from \cite{IceCube:2020wum}. We additionally use the pull method to account for uncertainties in the background and deposited-energy normalizations, with priors as in \cite{IceCube:2015gsk}. The results show that sizable changes in the NC normalization can significantly modify the expected event rate for neutrinos traversing the Earth, and are excluded by the current data.

\begin{figure}[t]
	\centering
	\begin{tabular}{ccccc}
\includegraphics[width=0.5\textwidth]{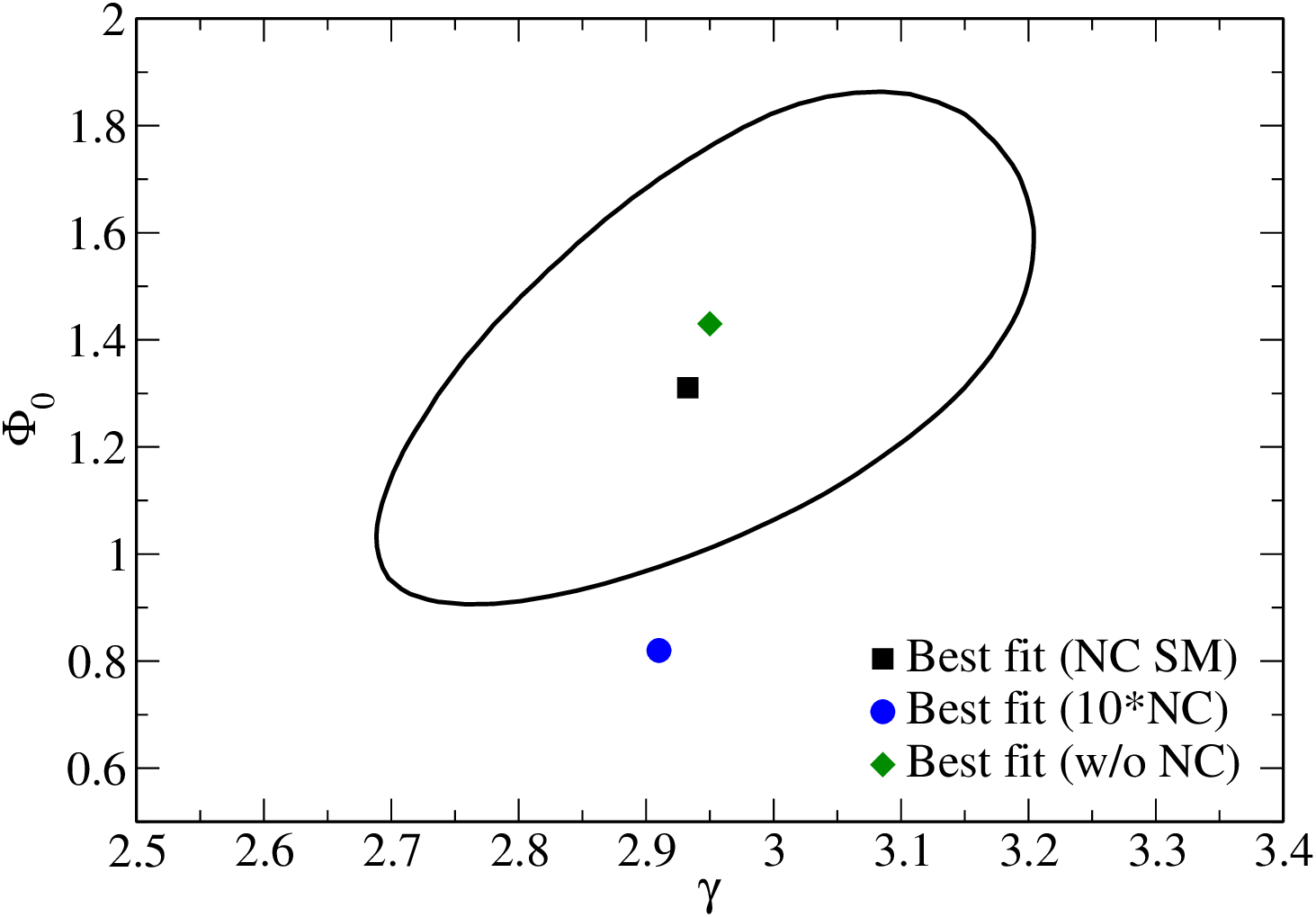}
	\end{tabular}
\caption{ Best fit for the astrophysical neutrino flux parameters obtained from likelihood analysis considering three scenarios for the normalization of the NC cross section and using 12 years of HESE data. For NC SM we also present the allowed region at 68\% CL. }
\label{fig:phi0_gamma}
\end{figure}

To estimate the sensitivity of the IceCube HESE data set to NC effects, we determine the best-fit neutrino-flux parameters (and corresponding priors) for several choices of the NC cross-section normalization. Figure~\ref{fig:phi0_gamma} shows the results for three scenarios: the SM NC prediction, an NC rescaling by a factor of 10, and the case with NC interactions switched off. For the SM case, we also show the 68\% confidence-level allowed region (solid black contour). We find that the spectral index $\gamma$ is only weakly affected by variations in the NC contribution, whereas increasing the NC strength leads to a decrease in the best-fit flux normalization.

\begin{figure}[t]
	\centering
	\begin{tabular}{ccccc}
\includegraphics[width=0.7\textwidth]{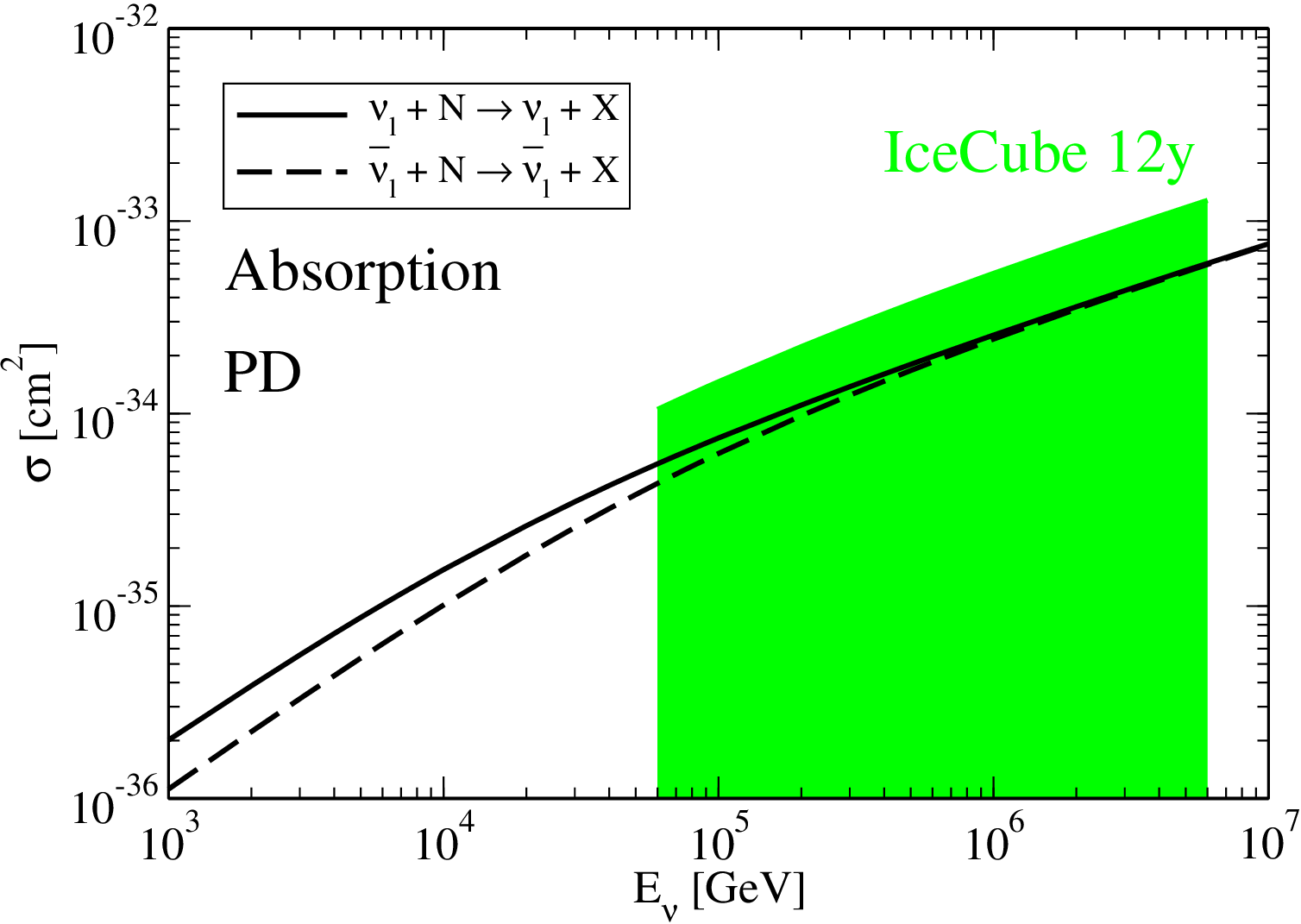}
	\end{tabular}
\caption{ Allowed region for the neutral-current  neutrino-nucleon cross section at the 68\% confidence level, constrained by Earth-absorption effects using 12 years of HESE data. For comparison, the Standard Model  predictions for neutrino and antineutrino cross sections are indicated by solid and dashed black lines, respectively. }
\label{fig:NC_absorption}
\end{figure}

In Fig.~\ref{fig:NC_absorption} we present the allowed NC cross section inferred from 12 years of IceCube HESE data using neutrino absorption in the Earth. We use a single deposited-energy bin (60~TeV--6~PeV) and allow an energy-independent rescaling of the NC cross section, while fixing the CC cross section to its SM prediction. At 68\% confidence level (CL), the data allow an enhancement of the NC normalization up to 2.14. With the current dataset, we do not obtain a meaningful lower bound on the NC cross section, and the best fit corresponds to vanishing NC interactions.

\subsection{Constraining the NC cross section using the excess of cascade events}
\label{sub:cascade}

In the previous subsection, we discussed how to constrain the NC neutrino--nucleon cross section through its impact on absorption and regeneration as neutrinos traverse the Earth. A complementary strategy is to study the effect of NC interactions on cascade-like events. For interactions that start within the instrumented volume, IceCube can identify three topologies: tracks, cascades, and double cascades. Here, we classify HESE events only as tracks or cascades. Track events contain a high-energy muon in the final state, typically produced in a $\nu_\mu$ (or $\bar\nu_\mu$) CC interaction or in the decay of a $\tau$ lepton; for simplicity, we include only $\nu_\mu$/$\bar\nu_\mu$ CC interactions and $\tau\to\mu$ decays. Cascade events are characterized by an approximately spherical energy deposition in the ice and can arise from $\nu_e$/$\bar\nu_e$ CC interactions, $\nu_\tau$/$\bar\nu_\tau$ CC interactions followed by $\tau$ decays to electrons or hadrons, or NC interactions of any flavor. 

\begin{figure}[t]
	\centering
	\begin{tabular}{cc}
\includegraphics[width=0.5\textwidth]{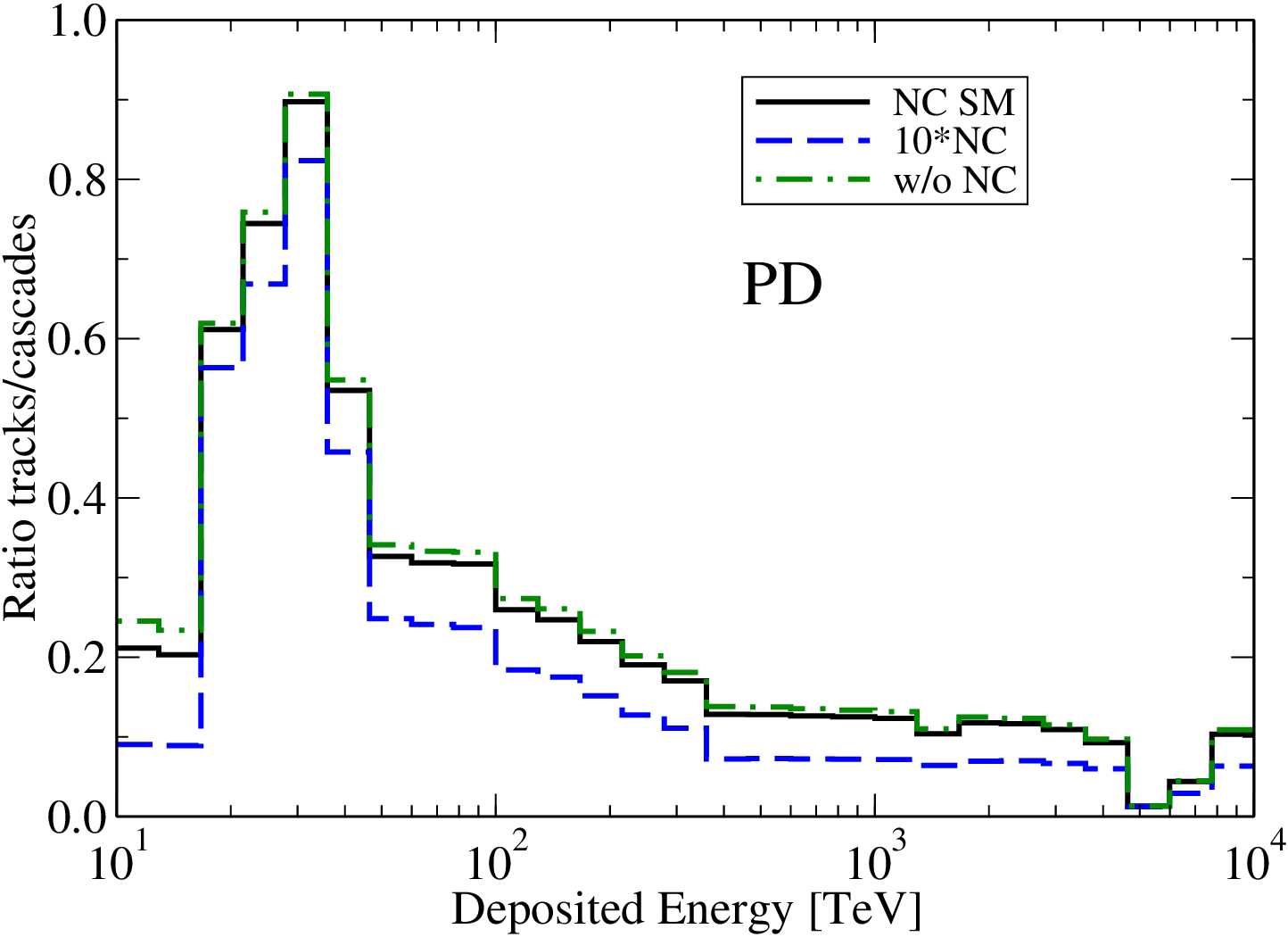} & 
\includegraphics[width=0.5\textwidth]{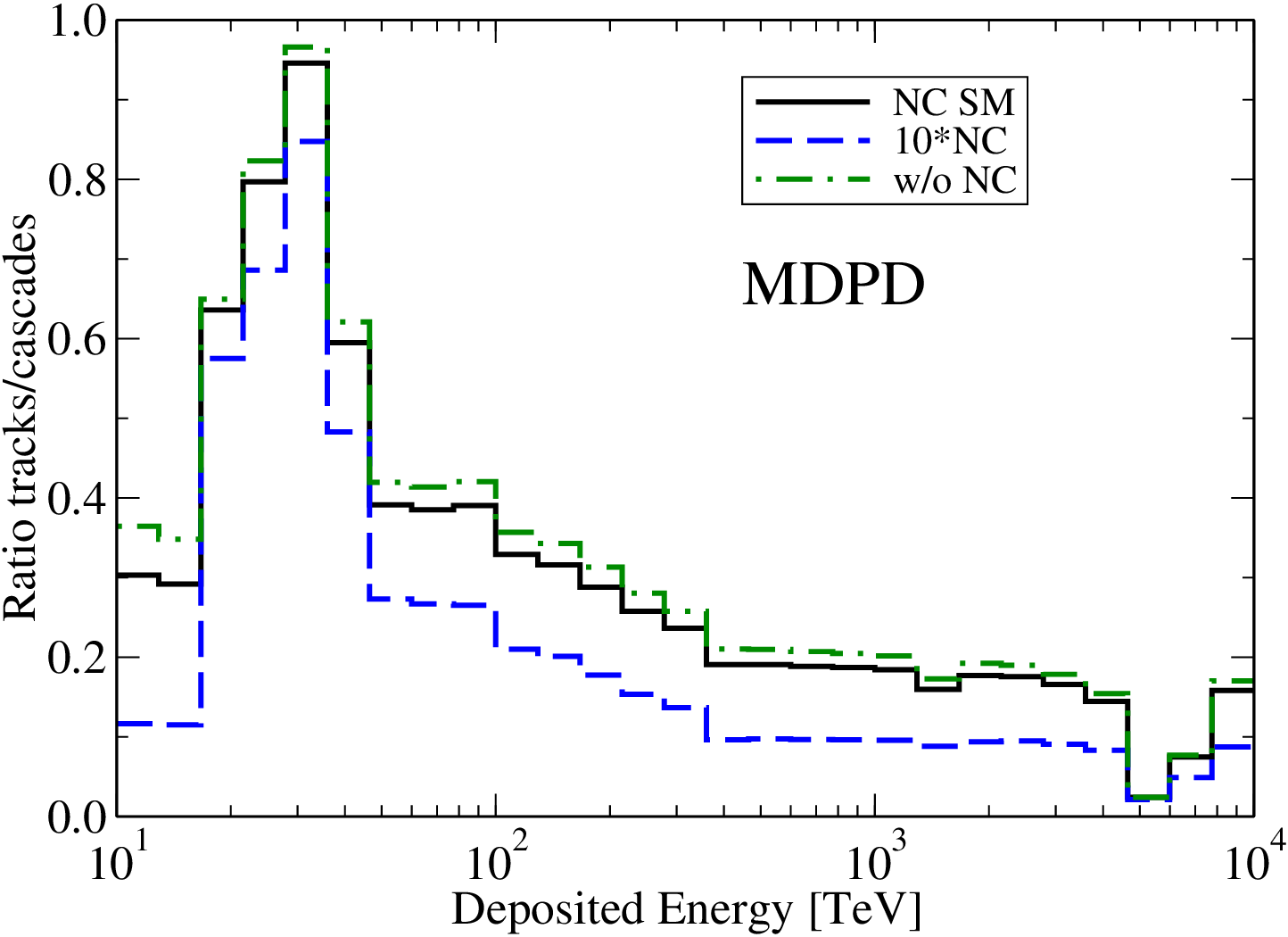}   
	\end{tabular}
\caption{Tracks-to-cascade ratio for IceCube HESE events as a function of deposited energy, assuming the astrophysical neutrino flavor compositions predicted by the PD (left) and MDPD (right) models. The results are shown for three illustrative cases of neutral-current cross-section normalization. }
\label{fig:ratio_tracks_cascades}
\end{figure}

In our analysis, we assume that the astrophysical neutrino flavor composition is known and consider two benchmark production scenarios: pion decay (PD) and muon-damped pion decay (MDPD). A more general fit in which the NC cross section and the flavor ratios are inferred simultaneously is beyond the scope of this work. To illustrate how the track and cascade yields depend on the assumed production scenario, Fig.~\ref{fig:ratio_tracks_cascades} shows the HESE track-to-cascade ratio as a function of deposited energy. We show this ratio for the  PD (left) and MDPD (right) scenarios, assuming different normalization factors for the NC cross section. Over the energy range relevant to our analysis, MDPD predicts an expected number of tracks that is about 30\% larger than in the PD case. In addition, a NC rescaling factor larger (smaller) than 1 decreases (increases) this ratio.

\begin{figure}[t]
	\centering
	\begin{tabular}{ccccc}
\includegraphics[width=0.5\textwidth]{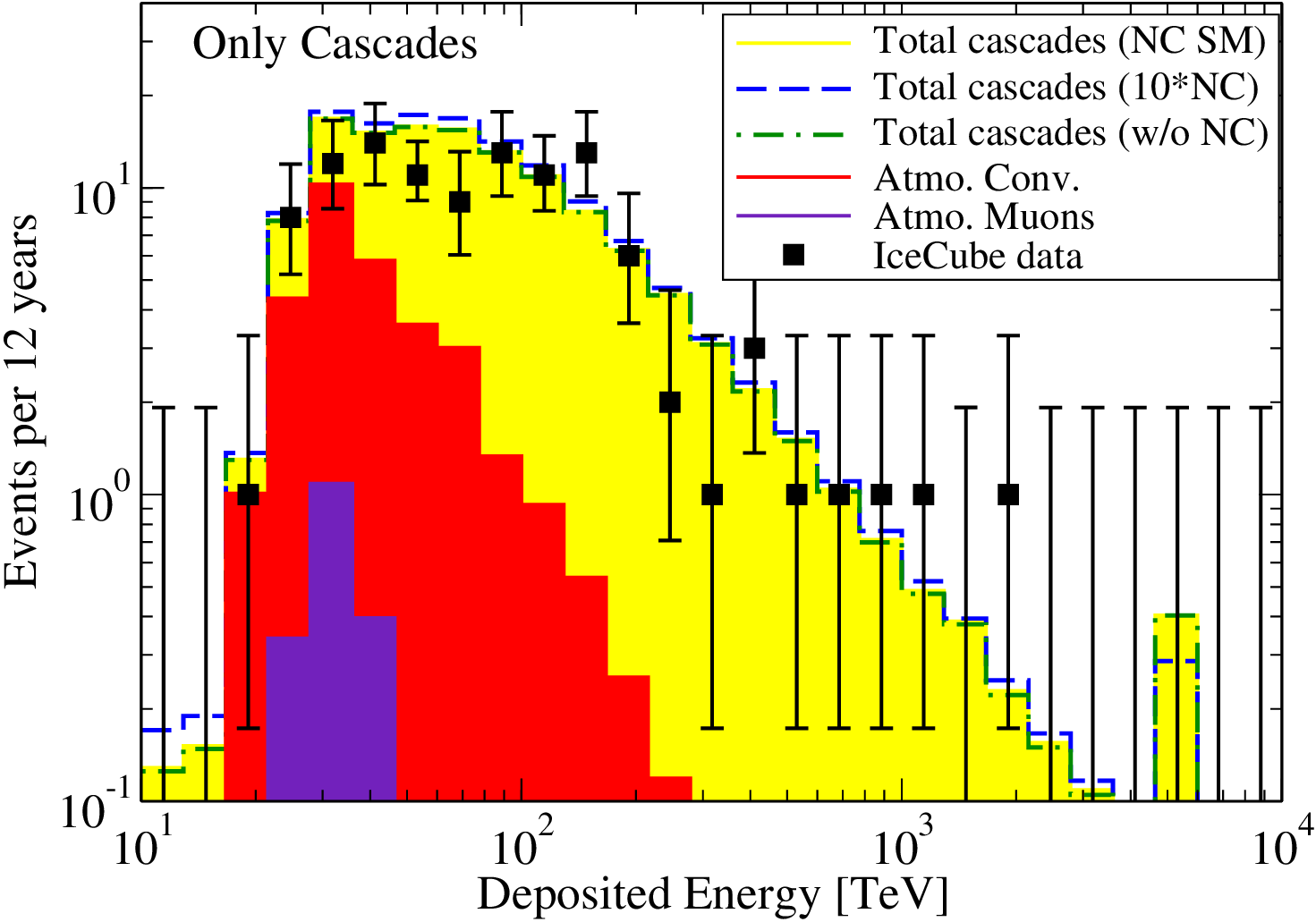}  &
\includegraphics[width=0.5\textwidth]{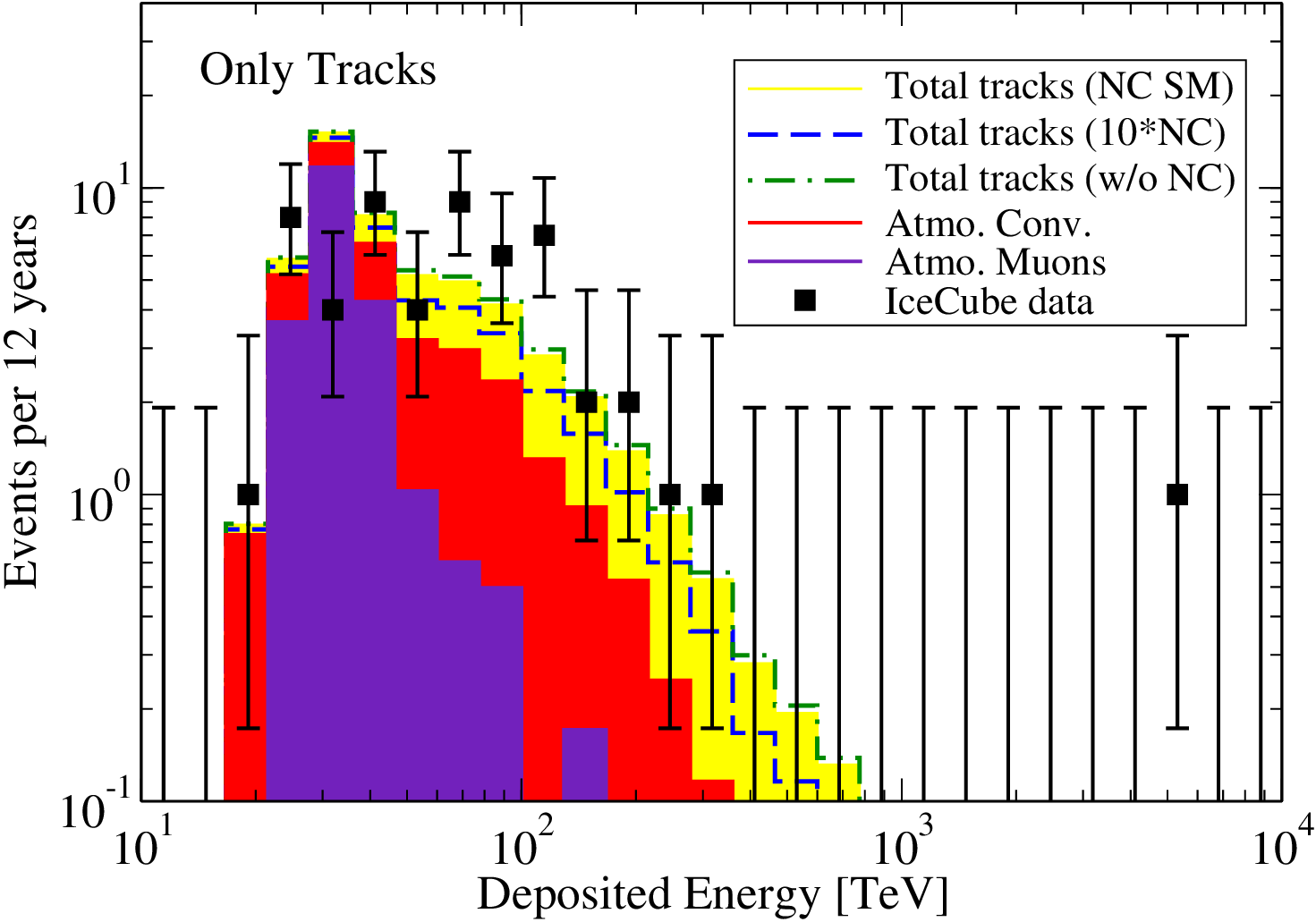}
	\end{tabular}
\caption{HESE cascade (left) and track (right) event distributions per bin of deposited energy, based on 12 years of data, comparing three distinct neutral-current  cross-section normalizations.}
\label{fig:events_cascades_tracks}
\end{figure}

\begin{figure}[t]
	\centering
	\begin{tabular}{ccccc}
\includegraphics[width=0.5\textwidth]{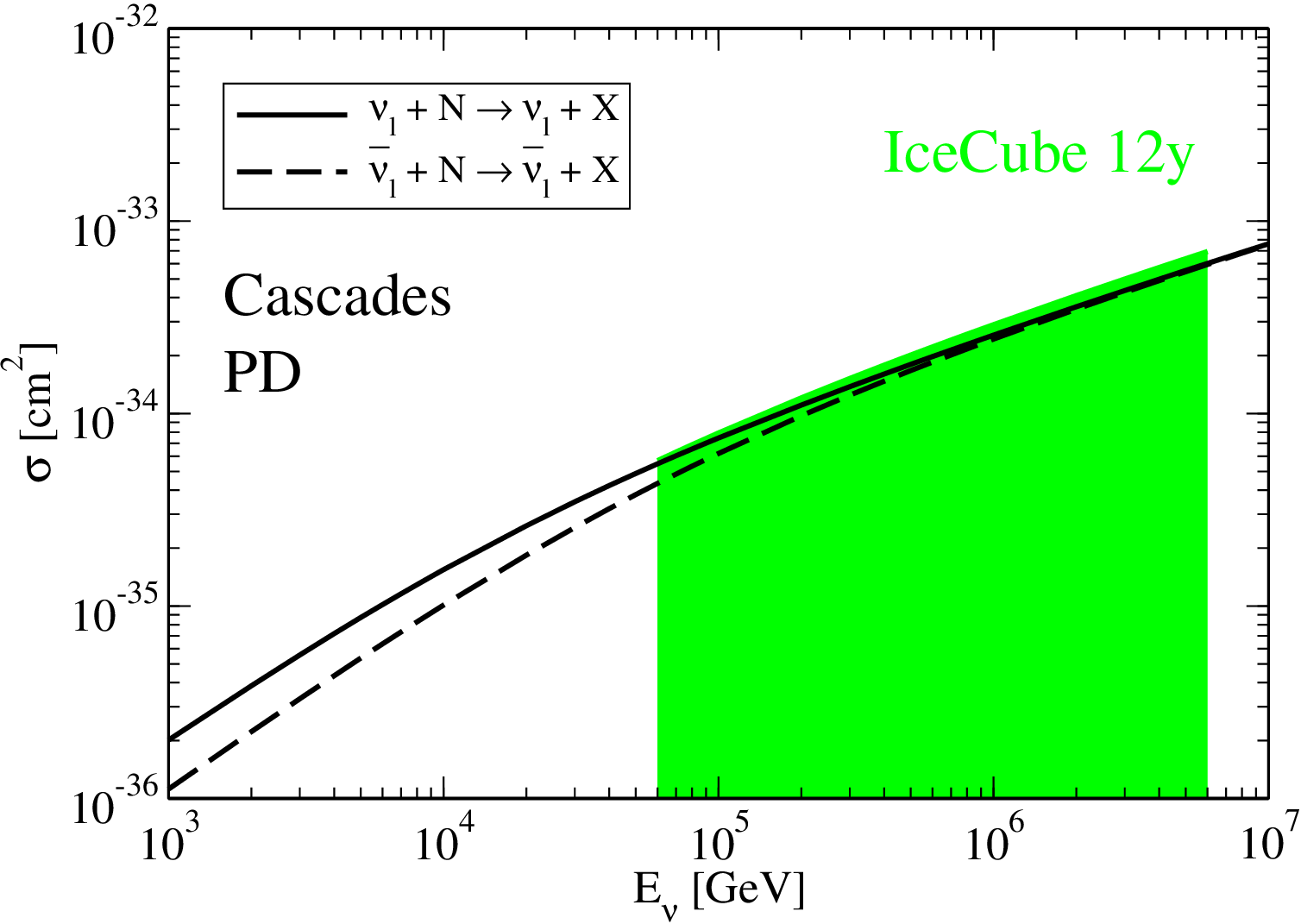} &
\includegraphics[width=0.5\textwidth]{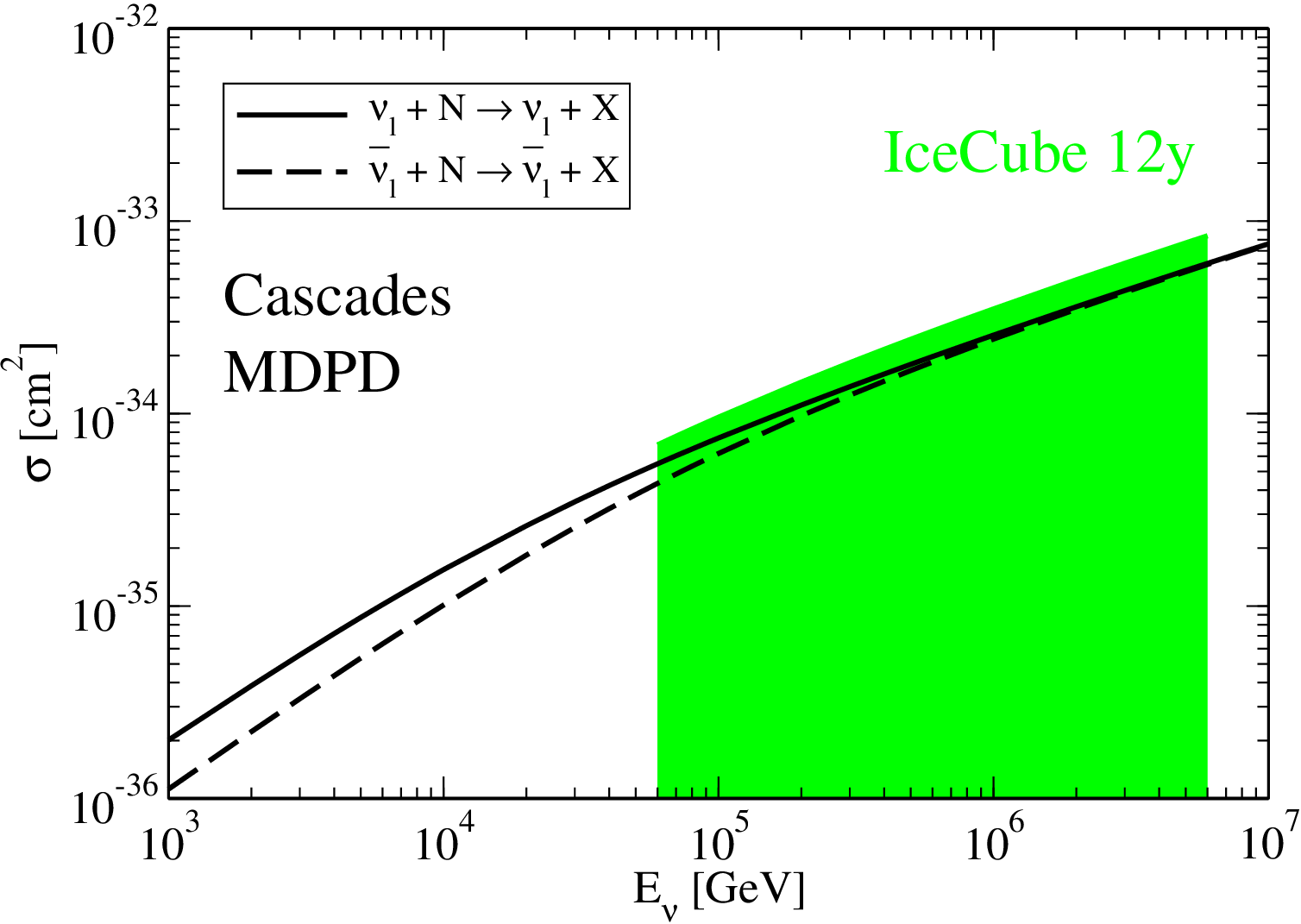}
	\end{tabular}
\caption{Allowed region for the neutral-current  neutrino-nucleon cross section at the 68\% confidence level, constrained by 12 years of HESE cascade data. The results correspond to neutrino flux models from pion decay (left) and muon-damped pion decay (right). Standard Model  predictions for neutrino and antineutrino cross sections are indicated by solid and dashed black lines, respectively. }
\label{fig:NC_cascades}
\end{figure}

We first analyze the PD scenario, which assumes an approximately equal flavor composition at Earth. Utilizing the 12-year HESE sample of tracks and cascades, we calculate best-fit flux parameters for various NC cross-section normalizations. Figure \ref{fig:events_cascades_tracks} displays the resulting deposited-energy distributions for cascades and tracks for these scenarios. 
In Fig.~\ref{fig:NC_cascades} (left) we present the 68\% CL constraint on the NC cross section obtained from the cascade analysis. The limit is derived from a likelihood computed separately for the cascade and track samples in Fig.~\ref{fig:events_cascades_tracks}, using only bins with deposited energy $E>60\,\mathrm{TeV}$. We obtain an upper bound of 1.16 times the SM NC cross section. This relatively strong constraint is driven by an excess of tracks relative to the expectation for an equal flavor composition at Earth.

The results associated with the astrophysical flavor ratios predicted by the MDPD scenario are presented in Fig. \ref{fig:NC_cascades} (right). In this case, we obtain a 68\% CL upper bound of 1.40 times the SM prediction, which is less stringent than the result obtained under the PD scenario. This comparison highlights that NC cross-section constraints derived from IceCube data are sensitive to the assumed neutrino flavor composition.

\subsection{Optimistic approach with a fixed astrophysical neutrino flux}
\label{sub:optimistic}

In the preceding subsections, we performed a comprehensive analysis, determining best-fit flux parameters and profiling nuisance parameters via the pull method for each choice of the NC cross section rescaling parameter. Here, we consider an optimistic scenario in which neutrino-flux parameters and nuisance-parameter priors are fixed by independent observations, thereby eliminating them as sources of uncertainty in the IceCube analysis. Specifically, we assume that the normalization and spectral index are fixed at $\Phi_0=1.31$ and $\gamma=2.93$, values obtained in subsection \ref{sub:absorption} for the SM NC prediction.

\begin{figure}[t]
	\centering
	\begin{tabular}{ccccc}
\includegraphics[width=0.7\textwidth]{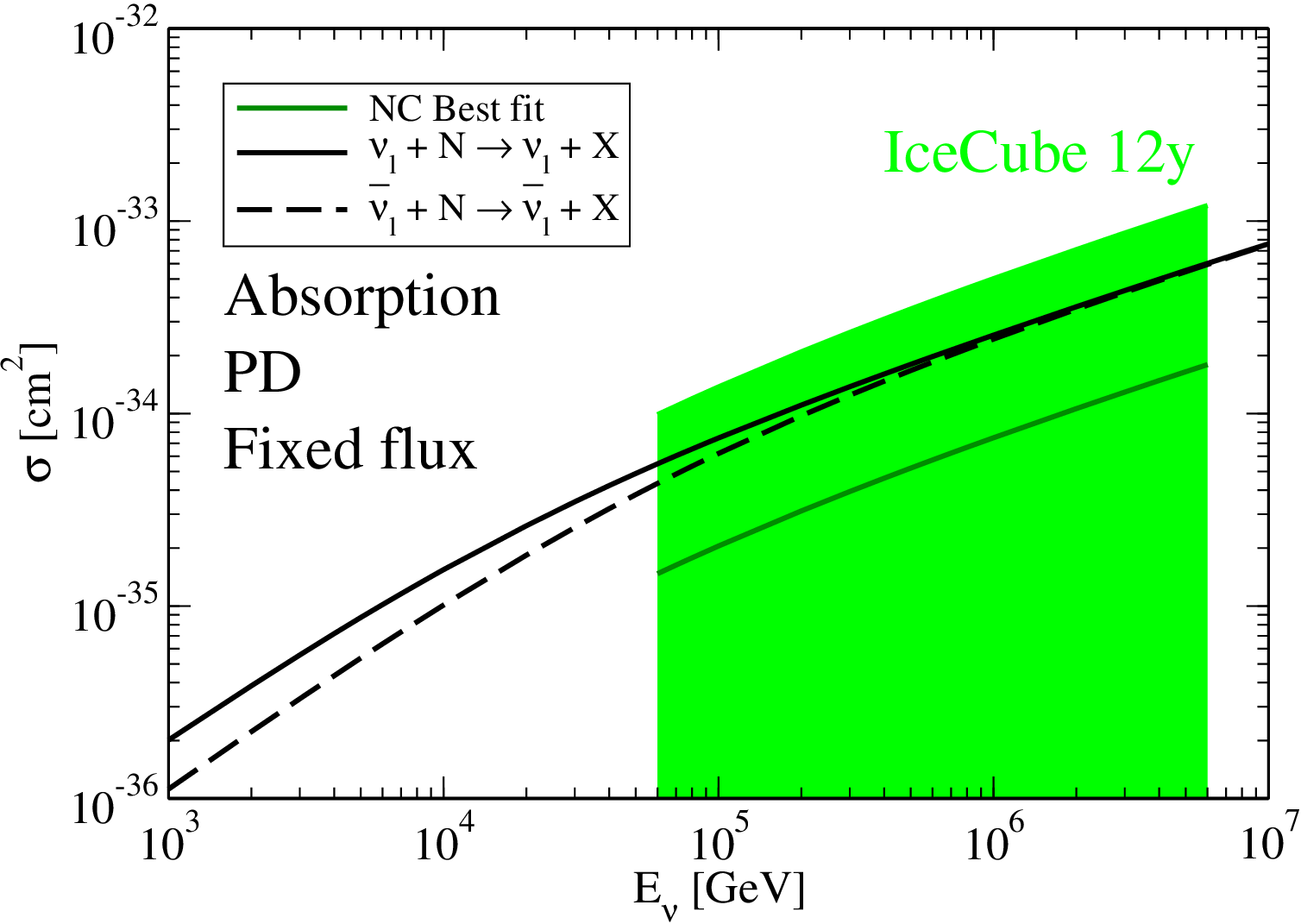}
	\end{tabular}
\caption{Allowed region for the neutral-current  neutrino-nucleon cross section at the 68\% confidence level, constrained by Earth-absorption effects using 12 years of HESE data. Standard Model  predictions for neutrino and antineutrino cross sections are indicated by solid and dashed black lines, respectively. This analysis assumes a fixed neutrino flux, with parameters corresponding to the best-fit values obtained under the SM NC assumption. }
\label{fig:NC_absorption_FF}
\end{figure}

Figure~\ref{fig:NC_absorption_FF} shows the allowed NC cross section inferred from 12 years of IceCube HESE data using neutrino absorption in the Earth, now fixing the astrophysical flux parameters. As before, we use a single deposited-energy bin from $60\,\mathrm{TeV}$ to $6\,\mathrm{PeV}$ and allow for an energy-independent rescaling of the NC cross section. At 68\% CL, the data allow an enhancement of the NC normalization up to 2.01, yielding a stronger limit than in the analysis with a fitted astrophysical flux. Even in this optimistic setup, the data do not set a lower bound on the NC cross section; the best fit is obtained for an NC cross section equal to 0.3 times the SM prediction.

\begin{figure}[t]
	\centering
	\begin{tabular}{ccccc}
\includegraphics[width=0.5\textwidth]{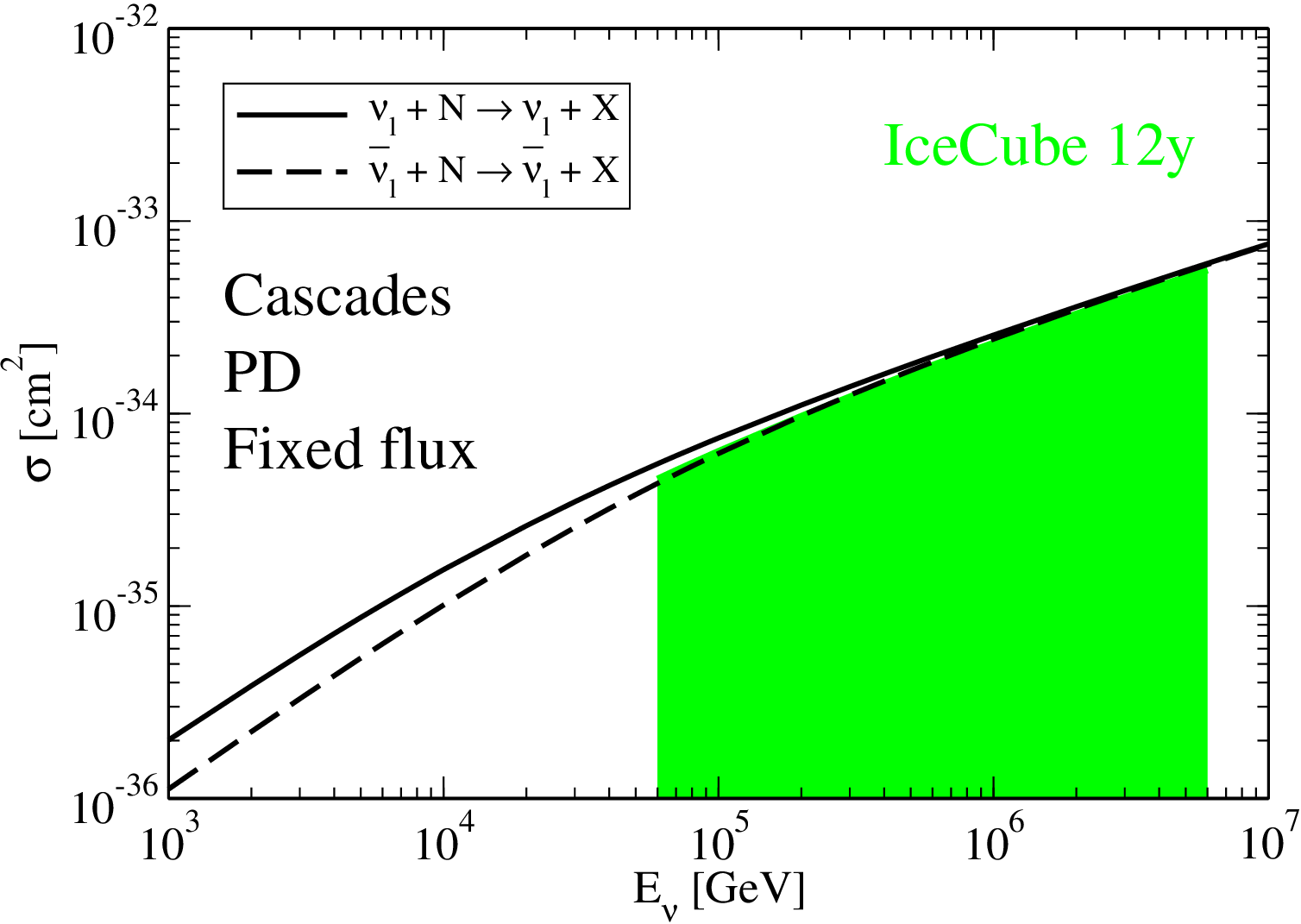} &
\includegraphics[width=0.5\textwidth]{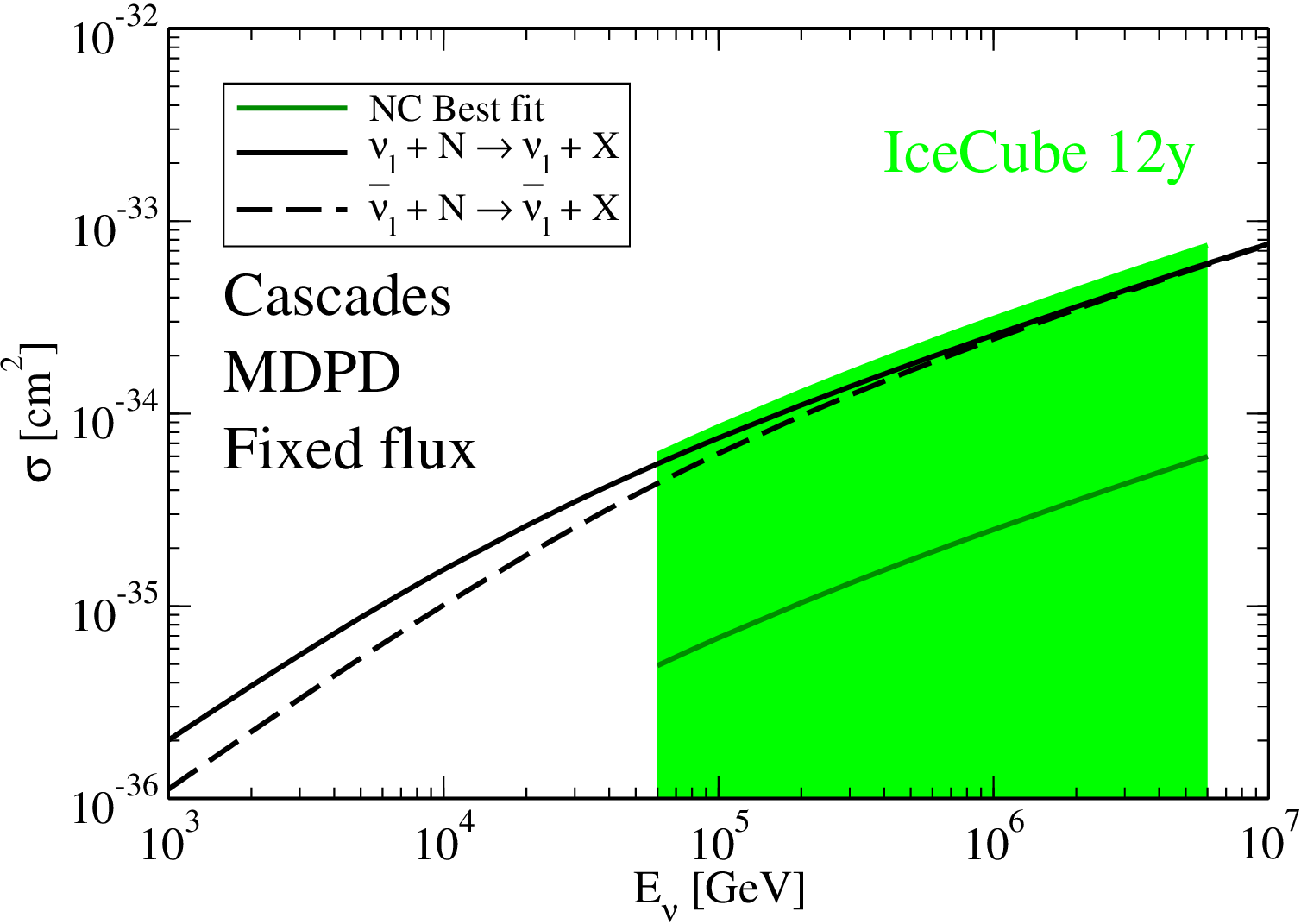}
	\end{tabular}
\caption{ Allowed region for the neutral-current (NC) neutrino-nucleon cross section at the 68\% confidence level, constrained by 12 years of HESE cascade data. The panels present neutrino flux models from pion decay (left) and muon-damped pion decay (right). For comparison, Standard Model  predictions for neutrino and antineutrino cross sections are indicated by solid and dashed black lines, respectively. This analysis assumes a fixed neutrino flux, with parameters corresponding to the best-fit values of the central SM NC prediction. }
\label{fig:NC_cascades_FF}
\end{figure}

Figure~\ref{fig:NC_cascades_FF} shows the results for the cascade analysis,  as in Fig.~\ref{fig:NC_cascades}, but now assuming that the astrophysical neutrino flux is known. In this case the constraints are stronger: the 68\% CL upper bounds are $\approx 1.0$ and 1.25 times the SM prediction for the pion-decay (left) and muon-damped pion-decay (right) scenarios, respectively. For the MDPD model, the best fit corresponds to an NC cross section of about 0.1 times the SM prediction. We emphasize that these limits are obtained under specific assumptions about the flavor composition at Earth and depend sensitively on those assumptions.

\subsection{Prospects for IceCube-Gen2}
\label{sub:gen2}

\begin{figure}[t]
	\centering
	\begin{tabular}{ccccc}
\includegraphics[width=0.5\textwidth]{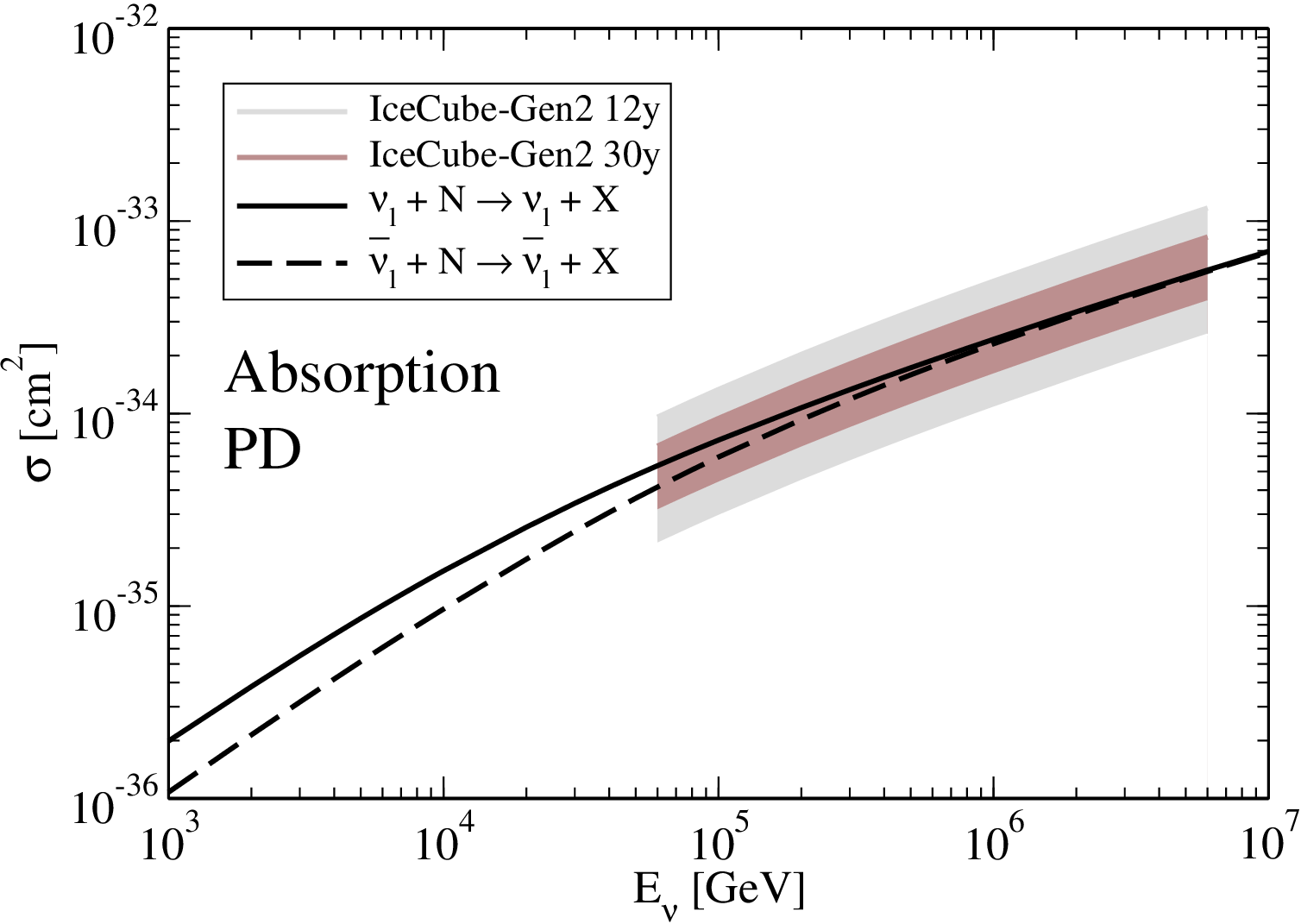}  &
\includegraphics[width=0.5\textwidth]{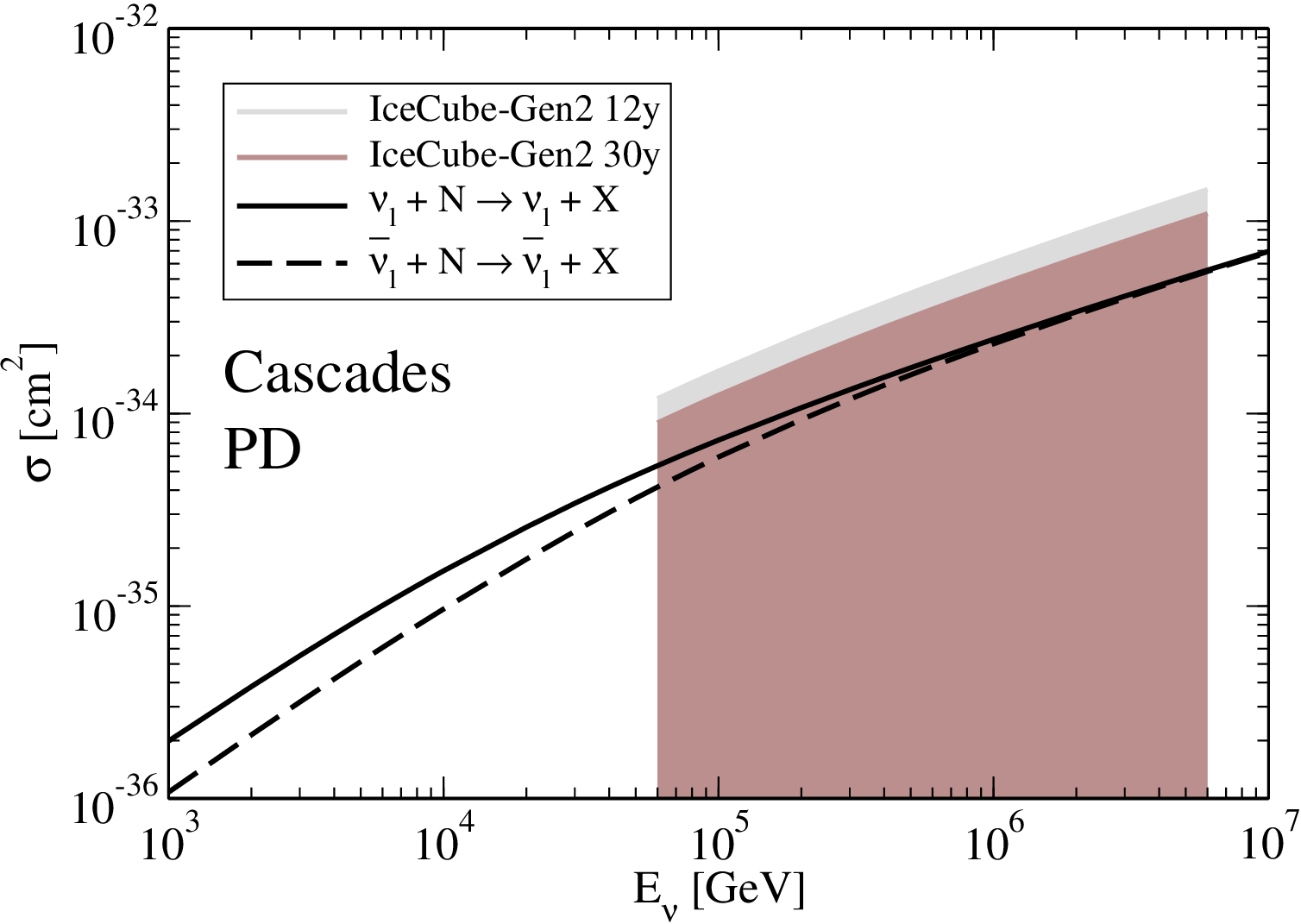}
	\end{tabular}
\caption{Projections for the 68\% CL allowed region of the neutral-current  neutrino-nucleon cross section at IceCube-Gen2, using 12 and 30 years of HESE data. Results are constrained by Earth-absorption effects (left) and cascade event data (right), assuming a pion-decay neutrino flux model. Standard Model  predictions for neutrino and antineutrino cross sections are indicated by solid and dashed black lines, respectively. }
\label{fig:NC_gen2}
\end{figure}

In this subsection, we present our projections for IceCube-Gen2. Figure~\ref{fig:NC_gen2} shows the expected IceCube-Gen2 sensitivity to the NC cross section using neutrino absorption (left) and the cascade sample (right), assuming a fitted astrophysical neutrino flux. We consider HESE exposures of 12 and 30 years. We find that the absorption-based analysis is expected to provide the stronger sensitivity. Nevertheless, the two approaches are complementary: the absorption method is limited primarily by uncertainties in the CC cross section but is only weakly dependent on the neutrino flavor composition, whereas the cascade-based method is largely insensitive to CC cross-section uncertainties (which affect tracks and cascades similarly) but depends strongly on the assumed flavor composition.

\begin{figure}[t]
	\centering
	\begin{tabular}{ccccc}
\includegraphics[width=0.5\textwidth]{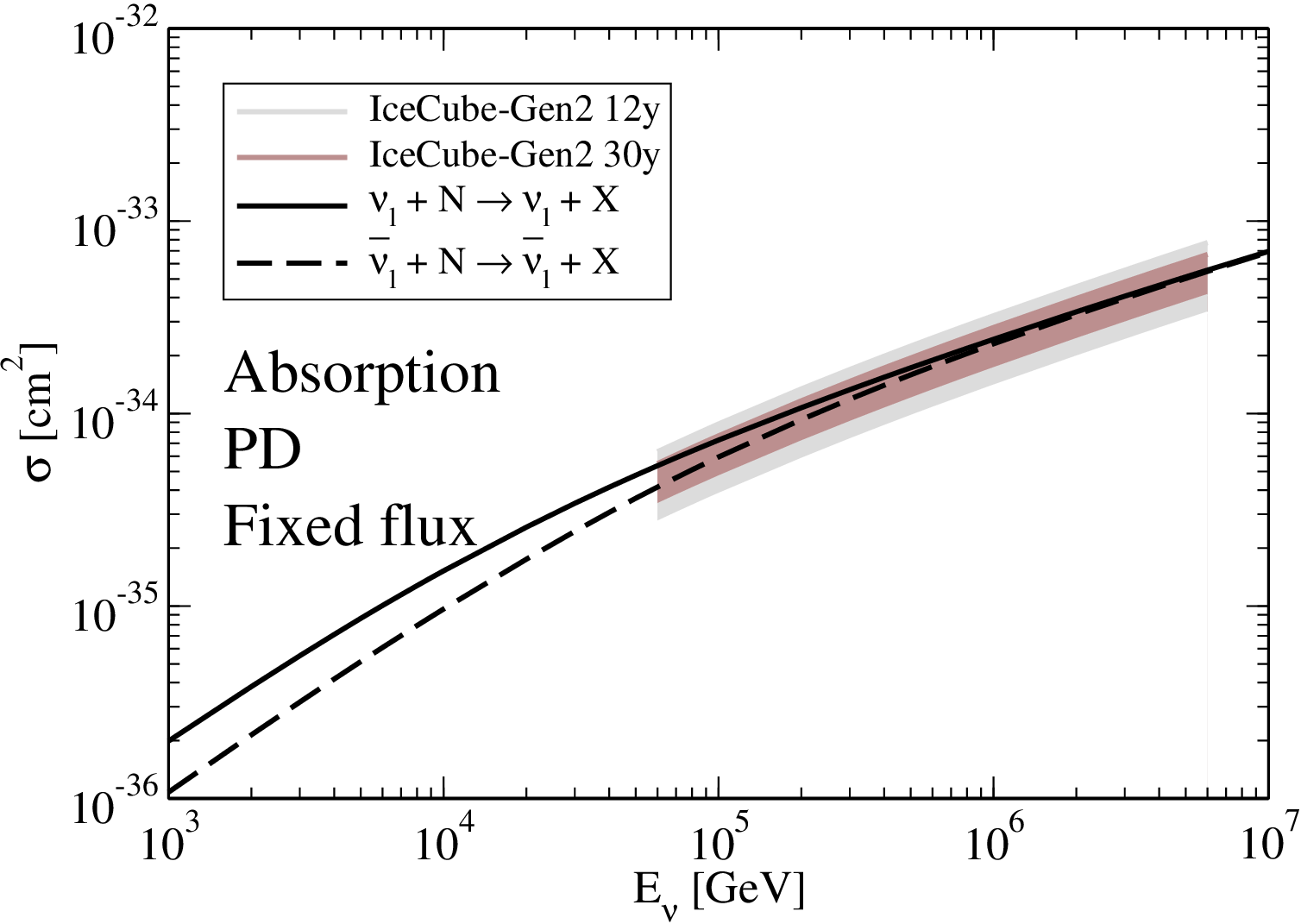}  &
\includegraphics[width=0.5\textwidth]{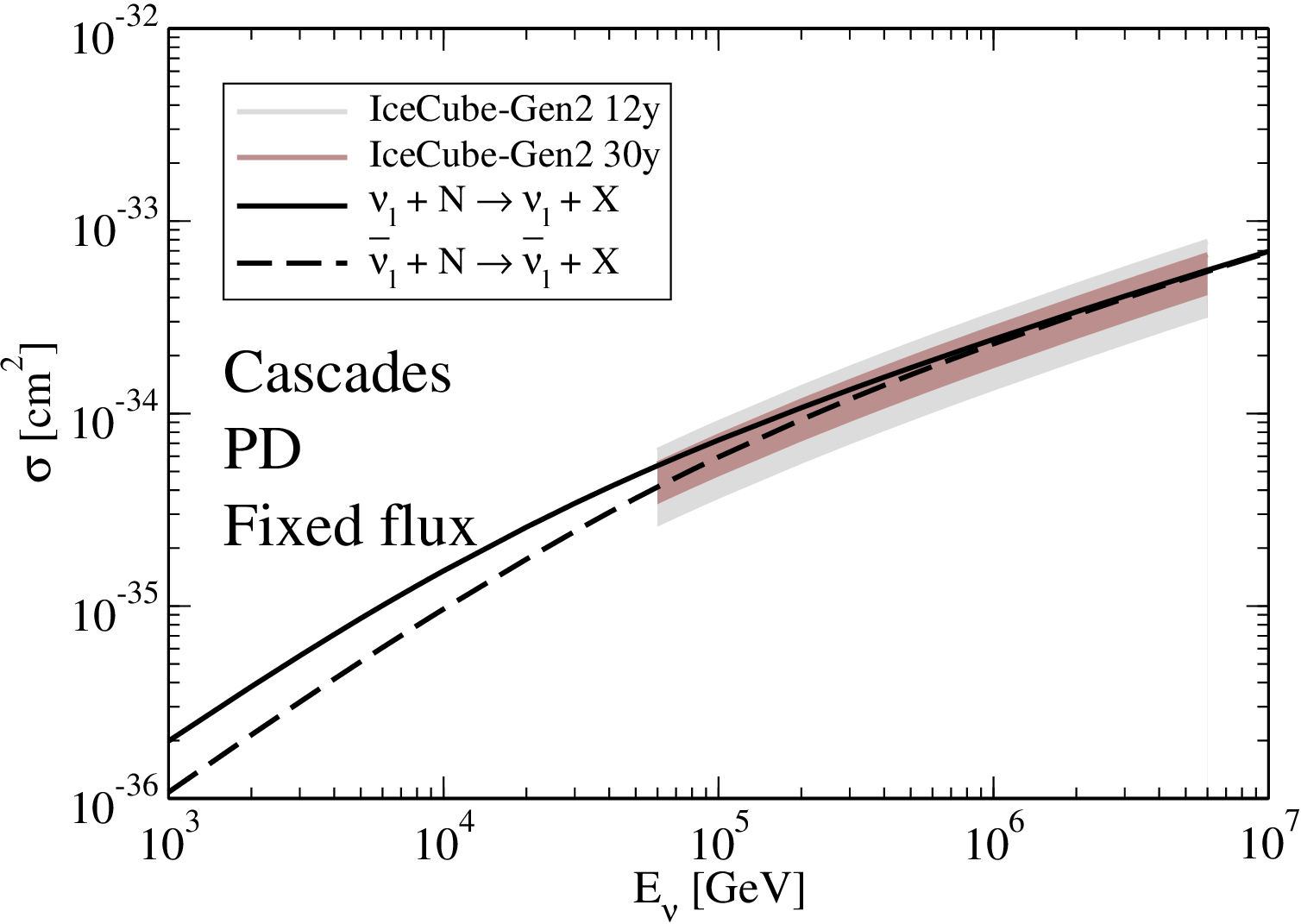}
	\end{tabular}
\caption{Projections for the 68\% CL allowed region of the neutral-current  neutrino-nucleon cross section at IceCube-Gen2, derived from 12 and 30 years of HESE data. Constraints are obtained from Earth-absorption effects (left) and cascade event data (right), assuming a pion-decay neutrino flux model. Standard Model  predictions for neutrino and antineutrino cross sections are indicated by solid and dashed black lines, respectively. This analysis assumes a fixed neutrino flux, with parameters set to the best-fit values of the  SM NC prediction. }
\label{fig:NC_gen2_FF}
\end{figure}

Finally, in Fig~\ref{fig:NC_gen2_FF} we present the same results as in  Fig~\ref{fig:NC_gen2}, for a fixed neutrino flux. In this scenario, the constraints on the NC cross section are significantly tighter. Similar outcomes are observed when considering predictions derived from neutrino absorption and cascade event rates. 

\section{Summary}
\label{sec:sum}

Neutrino deep inelastic scattering (DIS) has been measured by various experiments over recent years. However, the difficulty of identifying neutral-current (NC) DIS in event-by-event analyses, combined with its relatively low impact on total event rates, makes the NC cross section a challenging observable to extract. In this work, we investigate the potential for inferring the NC neutrino-nucleon cross section using IceCube data through two complementary approaches. First, we examine the impact of the NC cross section on neutrino absorption and regeneration as neutrinos traverse the Earth before reaching the IceCube Observatory. Second, we evaluate the feasibility of inferring the NC cross section from observed cascade and track event rates. Our results indicate that both approaches allow for the derivation of an upper limit on the NC cross section, although current data remain insufficient to establish a lower limit. We argue that these methods are complementary: the first is limited by uncertainties in the charged-current (CC) cross section and the assumption of astrophysical neutrino flux isotropy, while the second depends strongly on assumptions regarding neutrino flavor ratios at Earth. Additionally, we consider an optimistic scenario in which the astrophysical neutrino flux is fixed, demonstrating that more stringent constraints can be derived. Finally, we estimate the sensitivity of IceCube-Gen2; our results suggest that this proposed observatory will be capable of robustly constraining the NC cross section, including the establishment of a lower limit.

\begin{acknowledgments}
\noindent The work of R.F. was supported by FAPERGS. VPG was partially financed by the Brazilian funding agencies CNPq, FAPERGS and INCT-FNA  (Process No. 408419/2024-5). 
\end{acknowledgments}

\bibliographystyle{utphys}
\bibliography{refs}

\end{document}